\documentclass{aa}  
\usepackage{natbib}
\usepackage{amssymb}
\usepackage[colorlinks=true,linkcolor=blue,citecolor=blue,urlcolor=blue]{hyperref}
\usepackage{graphicx}
\usepackage[dvipsnames]{xcolor}
\usepackage{txfonts}
\usepackage{booktabs}
\usepackage{multirow} 
\usepackage{subcaption}
\usepackage{threeparttable}

\usepackage[normalem]{ulem}

\defcitealias{Bilicki2018}{B18}
\defcitealias{Bilicki2021}{B21}

\definecolor{mypink3}{cmyk}{0, 0.7808, 0.4429, 0.1412}
\definecolor{mypink2}{cmyk}{0, 0.8808, 0.0, 0.1412}

\definecolor{A}{rgb}{0.8, 0.2, 0.0}

\definecolor{B}{rgb}{0.5, 0.2, 0.4}

\begin{document} 
\date{}
\title{Hybrid-z: Enhancing the Kilo-Degree Survey bright galaxy sample\\photometric redshifts with deep learning}

\titlerunning{KiDS-Bright sample photo-$z$s with deep learning}

\authorrunning{John William, Jalan, Bilicki, et al.} 

\author{Anjitha John William\inst{1} \and Priyanka Jalan\inst{1} \and Maciej Bilicki\inst{1} \and Wojciech A. Hellwing\inst{1} \and \\ Hareesh Thuruthipilly\inst{2} \and Szymon J. Nakoneczny\inst{3}}

\institute{Center for Theoretical Physics, Polish Academy of Sciences, al. Lotnik\'{o}w 32/46, 02-668 Warsaw, Poland\thanks{\email{(anjithajm,pjalan,bilicki)@cft.edu.pl}}
\and National Centre for Nuclear Research (NCBJ),  ul. Pasteura 7, 02-093 Warsaw, Poland
\and Division of Physics, Mathematics and Astronomy, California Institute of Technology, 1200 E California Blvd, Pasadena, CA 91125, USA}

\abstract
{
We employed deep learning to improve the photometric redshifts (photo-$z$s) in the Kilo-Degree Survey Data Release 4 bright galaxy sample (KiDS-DR4 Bright). This dataset, used as foreground for KiDS lensing and clustering studies, is flux-limited to $r<20$ mag with mean $z=0.23$ and covers 1000 deg$^2$. Its photo-$z$s were previously derived with artificial neural networks from the ANNz2 package trained on the Galaxy And Mass Assembly (GAMA) spectroscopy. Here, we considerably improve on these previous redshift estimations by building a deep learning model, Hybrid-z, that combines an inception-based convolutional neural network operating on four-band KiDS images with an artificial neural network using nine-band magnitudes from KiDS+VIKING. The Hybrid-z framework provides state-of-the-art photo-$z$s for KiDS-Bright with negligible mean residuals of O($10^{-4}$) and scatter at a level of $0.014(1+z)$ -- representing a reduction of 20\% compared to the previous nine-band derivations with ANNz2. Our photo-$z$s are robust and stable independently of galaxy magnitude, redshift, and color. In fact, for blue galaxies, which typically have more pronounced morphological features, Hybrid-z provides a larger improvement over ANNz2 than for red galaxies. We checked our photo-$z$ model performance on test data drawn from GAMA as well as from other KiDS-overlapping wide-angle spectroscopic surveys, namely SDSS, 2dFLenS, and 2dFGRS. We found stable behavior and consistent improvement over ANNz2 throughout. Finally, we applied Hybrid-z trained on GAMA to the entire KiDS-Bright DR4 sample of 1.2 million galaxies. For these final predictions, we designed a method of smoothing the input redshift distribution of the training set in order to avoid propagation of features present in GAMA related to its small sky area and large-scale structure imprint in its fields. Our work paves the way toward the best-possible photo-$z$s achievable with machine learning for any galaxy type for both the final KiDS-Bright DR5 data and for future deeper imaging, such as from the Legacy Survey of Space and Time.
}

\maketitle

\section{Introduction}

Redshift is a key quantity for cosmological analyses. As the basic proxy for galaxy distances, it allows one to map the large-scale structure of the Universe in time and three-dimensional space. 
Redshift can be measured to a sub-percent accuracy only via spectroscopy. For such {spectroscopic redshifts} (spec-$z$s), 
one first collects the spectrum of an object and then identifies the shift in spectral lines with respect to the rest frame. However, even in the current era of fast measurements, such as with the Dark Energy Spectroscopic Instrument \citep{DESI2016}, spec-$z$s can be obtained only for a small fraction of all detected galaxies.

On the other hand, redshifts can be estimated for a much larger sample from photometric measurements. Such photometry-based redshifts (photo-$z$) provide an alternative method to the spec-$z$s and are based on the correlation between redshift and apparent galaxy magnitudes. This approach was originally pointed out by \cite{Baum1957} and first applied to obtain photoelectric magnitudes in nine passbands by \cite{Baum1962}. The two main attractive factors of photo-$z$s are that they allow one to obtain redshifts for galaxies fainter than generally possible with spectroscopy and that the number of objects with redshift estimates per unit telescope time is also much larger \citep{Hildebrandt2010}. Notably, photo-$z$s cannot provide redshift accuracy and precision as good as spec-$z$s. Nevertheless, they are indispensable in today's massive imaging surveys cataloging millions and billions of galaxies.

Photo-$z$s are based on a complicated mapping from photometry to redshift space, and they are difficult to handle analytically since this mapping depends on observational, computational, and statistical factors. Photo-z estimation methods can be generally categorized into template fitting and empirical approaches. In the latter case, which is our focus in this paper, the relation between photometric quantities and redshift is very commonly found by machine learning (ML) algorithms, although we note that simpler approaches of using functional fitting have also been proposed \cite[e.g.,][]{Connolly1995, KroneMartins2014}.  

In supervised ML techniques, the algorithm derives the empirical relation between observed quantities and labels from appropriate training on labeled data.
Therefore, their main challenge and limitation lies in extrapolating the results beyond a representative training set. However, if such appropriate training data exist, ML methods can excel, and this has led to numerous approaches being proposed for photo-$z$s based on the newest developments in computer science. Some examples include support vector machines \citep{Wadadekar2005}, random forest \citep{Carliles2010,Li2021}, artificial neural networks \citep[ANNs; e.g.,][]{Tagliaferri2003, Collister2004, Oyaizu2008}, ensemble learning \citep{Cunha2022}, Gaussian processes  \citep{Way2006,Bonfield2010}, self-organizing maps \citep{Way2012}, k-nearest neighbors \citep{Graham2018}, mixture density networks \citep{Ansari2021}, and finally deep neural networks \citep[e.g.,][]{Hoyle2016, D'Isanto2018}. 

Among the various supervised ML techniques for photo-$z$ derivation,  deep learning (DL) has emerged as a particularly promising one. Deep learning makes it possible to entirely skip "higher-level" quantities such as galaxy magnitudes or sizes derived from photometric post-processing and build the ML model using multi-band imaging directly. In such frameworks, "deep" indicates that the models usually have compounded multi-layer architectures. Their usage for photo-$z$s was pioneered by \cite{Hoyle2016} and then studied by, for example, \cite{D'Isanto2018,Menou2019,pasquet2019,Dey2022,Henghes2022,Treyer2023} for Sloan Digital Sky Survey (SDSS) data; \cite{Schuldt2021} for Hyper-Suprime Cam (HSC); \cite{Rui2022} for the Kilo-Degree Survey (KiDS); and by \cite{Roster2024} for DESI Imaging. These papers have demonstrated that using images directly, or in combination with magnitudes {\citep[e.g.,][]{Rui2022,Jones2024,Roster2024}}, allows one to derive photo-$z$s of better performance than those based on tabular galaxy data such as magnitudes.

A particular realization of DL is convolutional neural networks (CNNs). They are a type of ANN suitable for computer vision problems, such as image feature detection or classification, and are inspired by the human vision system. Similar to real neurons, which receive input and pass electrochemical signals \citep{McCulloch1943}, artificial neurons are used in CNNs. The significance of CNNs is evidenced by their vast use in image recognition tasks \cite[e.g.,][]{LeCun1998}. These networks are appropriate for processing data that have grid-like topology, such as images, because of their local connectivity, parameter sharing, and translational invariance. We chose CNNs to extract the galaxy image patterns by detecting features such as edges, textures, and shapes, which are expected to improve photo-$z$ derivations compared to methods that do not employ such information.

In this paper, we present a photo-z estimation in which we incorporate both fluxes (magnitudes) and multichannel galaxy images by using DL techniques for KiDS \citep{deJong2013}. KiDS is a multiband imaging survey covering about 1350 deg$^2$ of the sky, of which we employ $\sim1000$ deg$^2$ from its fourth data release \citep{Kuijken2019}. The DL photo-$z$s within KiDS have already been studied in detail by \cite{Rui2022}, where various setups using both images and magnitudes were compared for a general selection of galaxies spanning $0<z \lesssim 3$. 
However, until now, such imaging-based photo-$z$ approaches have not been employed in KiDS in the regime where they are expected to bring the most improvement over "shallow" ML, namely, for relatively bright and well-resolved galaxies. Here, we fill this gap and 
focus only on the bright end of the KiDS data. 

We studied the performance of DL for photo-$z$s in the flux-limited "KiDS-Bright DR4 sample," which includes all the KiDS galaxies within the magnitude cut of $r<20$ mag \cite[][hereafter \citetalias{Bilicki2018, Bilicki2021}]{Bilicki2018,Bilicki2021}. This sample is particularly useful for such an analysis, as by design it is selected to mimic the spectroscopic Galaxy And Mass Assembly dataset \cite[GAMA;][]{Driver2011}. As GAMA is flux limited and highly complete spectroscopically, it constitutes a very well matched training set for empirical photo-$z$ models. In the DL context, this aspect has been taken advantage of in the recent work by \cite{Treyer2023}, where photo-$z$ derivation for a sample of SDSS galaxies at $r<20$ mag was presented. The previous KiDS analysis by \cite{Rui2022} at the low-redshift end using DL as well as the dedicated studies by \citetalias{Bilicki2018, Bilicki2021} with "shallow" ML and GAMA training gave state-of-the-art photo-z results for the respective selections in KiDS. Therefore, here we aim to build on and extend these previous successful endeavors. Among our goals is to check if the current KiDS-Bright DR4 sample redshift estimates could be further improved. This is relevant, for instance, in view of the forthcoming Legacy Survey of Space and Time \citep{LSST2009}, where high-resolution multi-wavelength imaging of a depth greater than in KiDS will be available for the entire southern sky. Improvements in photo-z accuracy and precision of foreground galaxies are important in this context, as they help minimize related systematics of photometric clustering and lensing analyses.

The default photo-$z$s in KiDS data releases are derived with the Bayesian photometric redshifts \cite[BPZ;][]{Benitez2000} template-fitting tool. In particular, this tool is used to bin the weak lensing sources in redshift shells, and their true redshift distribution is then calibrated with self-organizing maps and via the clustering redshift technique \citep{Hildebrandt2021}. However, several studies have demonstrated that for bright low-redshift galaxies, empirical photo-$z$ methods can outperform the default BPZ solution when selecting galaxy samples from KiDS \citep[e.g.,][]{Cavuoti2015, Bilicki2018, Vakili2019, Rui2022}. This is possible if relevant training or calibration data are available to build a model mapping the photometric space to redshift using ML, but also other approaches such as red-sequence fitting \citep{Rozo2016, Vakili2019, Vakili2023} can be used as well. For the KiDS-DR4 Bright galaxy sample \citepalias{Bilicki2021} we are concerned with in this paper, the redshifts were estimated using ANNs from the public package ANNz2 \citep{Sadeh2016}.  The ANNs used photometric quantities (magnitudes, colors) as input and were trained with spectroscopic redshifts from GAMA. A number of tests have shown that such photo-$z$s are statistically accurate and precise \citepalias[i.e., have low mean bias and scatter;][]{Bilicki2018, Bilicki2021} not only for the KiDS-DR4 Bright sample as a whole but also for the sub-populations such as red and blue galaxies. In particular, this was possible thanks to the already mentioned intentional very good match of the galaxy selection in the KiDS-Bright DR4 sample to the GAMA training set.

In this work, we extend the previous feature-based ML efforts to build a successful DL model that we call "Hybrid-z." The model integrates both images and features for KiDS-DR4 Bright photo-$z$ estimations. We used the same GAMA training set as in \citetalias{Bilicki2021} and constructed a photo-z model that is conceptually similar to one tested in \cite{Rui2022}. Namely, it combines a deep convolutional network, employing $ugri$ imaging, with an ordinary ANN that is fed by nine-band galaxy magnitudes. An analogous configuration was also studied by \cite{Henghes2022} for SDSS, and inspired by their results, we also use "inception" as our basic architecture for the DL part. 

This paper is organized in the following manner. In Sect.~\ref{Data} we describe the data used. Next, in Sect. ~\ref{methodology} we explain the basic concepts of CNNs and the special CNN architecture that we used, called inception. 
In Sect.~\ref{model}, we describe our Hybrid-z model to estimate the photo-$z$s and the statistics that we used to measure the performance of the network. Sect.~\ref{result} presents our results, and in Sect.~\ref{conclusion}, we conclude and discuss future prospects.

\section{Data}
\label{Data}

In this section, we discuss the data used in this study. We employed the KiDS-Bright DR4 sample images supplemented with photometry (i.e., magnitudes). The required training and testing data are labeled using the spectroscopic redshifts from the GAMA survey.

\subsection{KiDS images and photometry}
\label{KiDS images}
Kilo-degree survey is an optical wide-field imaging survey of the European Southern Observatory (ESO) at the Very Large Telescope (VLT) Survey \cite[VST,][]{Capaccioli2011}, having at the focal plane a 268 million pixel Charge-Coupled Device (CCD) mosaic camera called OmegaCAM \citep{Kuijken2008}. VST is an alt-az mounted modified Ritchey-Cretien telescope located in the ESO Paranal Observatory, Chile. The images were taken in four broad bands ($ugri$), and the survey covers 1350 square degrees of the extragalactic sky. The final footprint of the survey is shown in fig.~3 of \cite{Wright2024}; here we use its publicly available subset.

KiDS-ESO Data Release 4 (KiDS DR4; \citep{Kuijken2019}), is the fourth public release of KiDS. The ASTRO-WISE optical pipeline and data reduction environment \citep{McFarland2013} is used to produce stacked (or co-added) composite images created by combining multiple individual exposures of the same sky area, one in each of the four bands. 
As a result, KiDS DR4 optical data are organized into $4\times1006$ one square-degree tiles. 

KiDS DR4 products consist of astrometrically and photometrically calibrated co-added images with a uniform pixel scale of 0.2 arcsec. The pixel units are fluxes relative to the 0th magnitude \citep{deJong2015}. We have downloaded the KiDS DR4 tiles\footnote{\url{https://kids.strw.leidenuniv.nl/DR4}} and made cutouts of galaxies with a size of $7.2''\times7.2'' $ ($36\times36$ pixels), since most of the objects we use are smaller than this. In particular, this cutout size is above the 99-percentile level of the half-light diameter (i.e., 2 $\times$ \texttt{FLUX\_RADIUS}) in KiDS-Bright. The largest galaxies, not fitting within our cutouts, are likely of little interest for our work anyway: they will be very nearby and will have had spec-$z$ from wide-angle surveys or otherwise will not be useful for lensing studies.

We have also tried bigger cutouts such as $20.2''\times20.2''$ and $8''\times8''$, motivated receptively by \cite{Grespan2024} and \cite{Rui2022}, but our finally adopted size gave the best results. The smaller size reduces the noise in images without losing galaxy flux information. Too large cutouts could also lead to frequent situations when more than one galaxy appears in the image. This type of contamination could adversely affect the performance of the model, although it has been argued in the literature that CNNs for photo-z estimation could in fact benefit from physically close pairs in the images \citep[e.g.,][]{pasquet2019}. In our case we also use magnitudes of individual galaxies, which should mitigate this effect, be it positive or negative. In any case, for our fiducial cutout size, more than one object is present in the image very rarely, in less than 1\% cases.

Finally, we normalized the pixel values to the range $[0,1]$ galaxy-wise, i.e. jointly for all the $ugri$ bands for a given galaxy cutout. The normalization formula is
\begin{equation}
    X_{normalized}=\frac {X-X_{min}}
    {X_{max}-X_{min}}.
\end{equation}
The minimum value of the pixels for the 4-band per-galaxy images (cutouts) is denoted as $X_{min}$ and the maximum value as $X_{max}$. In occasional cases when the image has (a) saturated pixel(s), this normalization will not work properly. Dealing with this problem would be beyond the scope of our work as it would require building an extra framework to pre-analyze all the 4.8 million cutouts and cleaning them up of such artifacts. We note that in most cases, objects located in such corrupted areas will not be useful for science anyway, as they will bear an appropriate KiDS \texttt{MASK} value indicating that their photometry is not reliable (see next Section). According to our estimates, for the `clean' data (i.e., those not affected by the mask), the presence of artifacts in the images is very infrequent, at a fraction of a percent level. Last but not least, the usage of magnitudes together with images in the model, will minimize their influence on the derived photo-zs.

KiDS DR4 photometric data consists of optical $ugri$  and near-infrared (NIR) data from the VISTA Kilo-degree INfrared Galaxy survey \cite[VIKING,][]{Edge2013} with observations in five bands: $ZYJHK_s$. In KiDS the magnitudes are by default derived with the Gaussian Aperture And Point spread function \citep[GAAP,][]{Kuijken2008} methodology. GAAP magnitudes are meant to provide robust galaxy colors irrespective of PSF differences in various bands, which makes them optimal for photo-$z$ derivation. This is an important asset for general weak lensing tomographic studies \citep{Kuijken2015}. GAAP magnitudes were also shown to be optimal for low-redshift photo-$z$ estimates in KiDS, as compared to other galaxy magnitude measurements available in this survey \citepalias{Bilicki2018}. The GAAP magnitudes in KiDS are provided in the AB  
system and their zero-point calibration 
is achieved by using coadd overlaps and stellar locus regression. They are corrected for Galactic extinction \citep{Schlegel1998} $E(B-V)$ map with \citep{Schlafly_2011} coefficients. 

We standardized these 9-band magnitude features using the StandardScaler class from the {\tt scikit-learn} Python library \citep{Fabian2011}, which computes the mean ($\overline{m}$) and standard deviation ($\sigma_{m}$) for each band independently. The magnitude values ($m$) in each band are transformed as
\begin{equation}
    m_{standardized}=\frac {m-\overline{m}}{\sigma_{m}}.
\end{equation}

In comparison to the optical $ugri$ images from KiDS, "coadds" are not readily available for the VIKING NIR, as these data are not processed by the ASTRO-WISE pipeline. Because of this, here we employ only the 4-band optical images, leaving the possible extension with NIR imaging to future work.

\subsection{KiDS-DR4 Bright  galaxy sample}
\label{sec:kids-bright}

In this work, we derive DL photo-$z$s for the KiDS-DR4 Bright galaxy sample, introduced in \citetalias{Bilicki2021}. This dataset contains galaxies selected from KiDS DR4 with the flux limit $r_\mathrm{auto}<20$ mag, where "auto" stands for {\tt SExtractor}-derived estimate of the total flux via automatic aperture photometry \citep{Bertin1996}. The selection of the sample is designed to have the best possible match with overlapping GAMA-equatorial spectroscopic data, which originally were selected using SDSS Petrosian magnitudes in the $r$ band \citep{Liske2015}. As it was discussed in \citetalias{Bilicki2021}, among the various $r$-band magnitude measurements in KiDS, a cut in $r_\mathrm{auto}$ provides the best correspondence to the GAMA original $r_\mathrm{Petro}<19.8$ limit.

In addition to the flux limit, the KiDS-Bright dataset uses various flags from the KiDS input catalog to remove point sources (stars, quasars) and artifacts. These in particular were $\mathtt{CLASS\_STAR}<0.5$ \& $\mathtt{SG2DPHOT}=0$ \& $\mathtt{SG\_FLAG}=1$ to select extended sources and $\mathtt{IMAFLAGS\_ISO}=0$ \& $(\mathtt{MASK} \& 28668)>0$ to remove imaging artifacts. However, we note that the 
published KiDS-Bright catalog also includes the objects without this final criterion applied, and in Sec.~\ref{Test sample} we test how the masking affects our photo-$z$ derivation. 

\subsection{GAMA spectroscopic data}
\label{GAMA}

In the training and testing phase, we used KiDS galaxies that have counterparts in the GAMA catalog \citep{Driver2011}, giving us the true redshift labels. GAMA is a multi-wavelength and spectroscopic survey in five fields\footnote{See \url{https://www.astro.ljmu.ac.uk/~ikb/research/gama_fields/} for the GAMA field locations.} (three equatorial: G09, G12, and G15, and two southern ones: G02 and G23), with a total $\sim$286 deg$^2$ area. Spectra were collected using the AAOmega fiber-fed spectrograph facility on the 3.9-m Anglo-Australian Telescope. GAMA provides spectra, redshifts, their quality marks, and other ancillary information.

KiDS DR4 fully overlaps with four GAMA fields (all but G02). Of these, the equatorial ones present the largest flux-limited spectroscopic completeness, originally estimated as 98.5\% at $r_\mathrm{SDSS}<19.8$ mag \citep{Liske2015} but subsequently revised to 98\% at $r_\mathrm{KiDS}<19.58$ \citep{Driver2022} after ingestion of KiDS photometry into the GAMA database \citep{Bellstedt2020}. Following the results of \citetalias{Bilicki2021}, where the addition of the shallower and less complete G23 data did not lead to improvement in photo-$z$ estimates, also here we used only the equatorial fields to train the model, and it is tested on both equatorial and G23 fields.
Similarly, other datasets such as for instance SDSS or DESI Early Data Release, are not sufficiently complete at our flux limit of $r<20$ mag to provide useful training sets for the overall galaxy sample (see, e.g., \citealt{Jalan2024}). Some of them, however, are useful for a posteriori tests of our photo-z model, and this is discussed in Sec.~\ref{Sec: external spec-z}.

In this work, we employ the GAMA-II spectroscopic redshift catalog of galaxies from the final GAMA DR4 \citep{Driver2022}.\footnote{\url{http://www.gama-survey.org/dr4}} 
For secure redshifts, we select galaxies with a normalized quality parameter $ NQ\geq3 $ \cite[see][for details]{Liske2015} and $z > 0.001$. 
We have identified galaxies within KiDS tiles that are shared between the GAMA and KiDS-Bright samples based on their right ascension and declination coordinates, assuming a 1" matching radius between KiDS and GAMA. Then we made cutouts of size $7.2''\times7.2''$ from KiDS images with these galaxies positioned at the center.

\section{Methodology}
\label{methodology}
\subsection{Convolutional neural networks}
\label{CNN}

A CNN is a sequence of layers: convolutional, pooling, and fully connected layers. The convolution takes place between input images and a small matrix of weights called a kernel or filter. Initially, the weights are small random values. 
In a CNN, the learning is hierarchical. The first convolutional layers are responsible for extracting the lower-level features of images, such as edges, corners, and textures \citep{Goodfellow-et-al-2016}. The kernels adjust their weights to extract these features. The filters in the following layers help integrate more complex features of input data. The network is trained by non-linear optimization of weights and biases through a gradient-descent algorithm. Input data is transformed by convolution operation, producing linear activation. 

The relationship between input data and output labels is usually non-linear. The linear relations are limited for this complex mapping from input to output spaces. To introduce non-linearity, the convoluted output passes through a non-linear function in hidden and output layers. This 
decides which neurons should activate. Without using the activation function, the output would be linearly dependent on the input, since the convolution output is a linear combination of input pixels within the receptive field. Rectified Linear unit (ReLu) and its variants such as peaky ReLu and leaky ReLu, hyperbolic tangent, and sigmoid functions are the common activation functions \citep{Jentzen2023}. The output of CNN is multidimensional (a tensor) and referred to as a feature map; it is another representation of the input data. 

The next stage is the pooling operation to modify the output. The pooling function replaces the output of the network at a certain location with a summary statistic of the nearby outputs \citep{Goodfellow-et-al-2016}. We experimented with various pooling operations, including average pooling and max pooling \citep{Gholamalinezhad2020}. Based on evaluation metrics,  we found that average pooling 
enhances model performance, which might be thanks to the fact that it deals better with noise in the images than max pooling. Average pooling computes the average of a rectangular neighborhood.

\subsection{Training}
\label{Training}

The training of a neural network involves several key steps and considerations. Initially, the dataset is divided into three subsets {in a random manner using \texttt{scikit-learn} python library}: training, validation, and testing set, in our case in a ratio of 70:15:15. The validation set serves to monitor the performance of the network during the training process.{ Also, there is no selection bias in magnitudes or redshifts in any subset.} {The final cutout catalog contains $\sim173$k galaxies in the equatorial fields and  $\sim125$k of these are used to train the model. The model is validated on $\sim26$k galaxies in the equatorial fields, and the remaining data are used for testing. Finally, it is applied to the entire KiDS-DR4 Bright sample of about 1.2 million galaxies.}

During training, the network updates its kernel values based on feedback signals. This iterative process aims to minimize the loss function, which represents the discrepancy between the predicted values and the true value. In the case of a CNN, the loss function is dependent on the weights. The weights are multiplied by the input values during the forward pass of the network to produce the output.

To minimize the loss function and approach the global minimum, the weights need to be updated in the opposite direction of the gradient. This optimization process involves gradient-based techniques. The{learning rate}, a crucial hyperparameter, determines the size of the step taken along the negative gradient direction. A too-small learning rate can result in slow convergence, while a too-large one can lead to divergent behavior of loss function.

The training process iterates over the dataset in batches, where the number of iterations is referred to as epochs, and the number of samples per batch is the batch size. At each epoch, the network computes the gradients of the weights with respect to the loss on the batch and updates the weights \citep{chollet2017}.

\subsection{Data augmentation}
\label{data augmentation}

Optimization and generalization play important roles in ML problems. Optimization is the process of adjusting the model parameters to get the best performance on training data, while generalization determines the model performance on unseen data. 
If the ML model is too simple to capture the underlying structure of the data, it performs poorly not only on the training data but also on unseen data. This is because it fails to learn relevant patterns from the training, leading to underfitting. On the other hand, overfitting happens when a model learns not only the underlying patterns but also the noise and random fluctuations present in the training \citep{Xia2024}. 
As a result, it performs very well on the training data but poorly on unseen data because it has essentially memorized the training data instead of generalizing from it.

One of the solutions for over- and underfitting
is generating a large training dataset using data augmentation. Data augmentation is 
commonly used in ML, particularly in the context of image classification and computer vision tasks \citep{Shorten2019}. It involves generating additional training data by applying various transformations (rotation, width and height shift, flipping, etc.) to the existing training samples. After learning a certain pattern in an image, a CNN can recognize it anywhere due to its translational invariance. 

We experimented with various data augmentation approaches in the training sample images and selected those that were most efficient for our case. We applied 5\% shifts in both width and height. Also, we performed horizontal and vertical flipping on the images. By these four variations, we extended our training set to $\sim484$k objects and were able to considerably improve model performance. We also tried image rotations, but these did not lead to any improvements. 

\subsection{Inception module}
\label{inception}

\begin{figure}
  \centering
  \includegraphics[width=0.95\columnwidth]{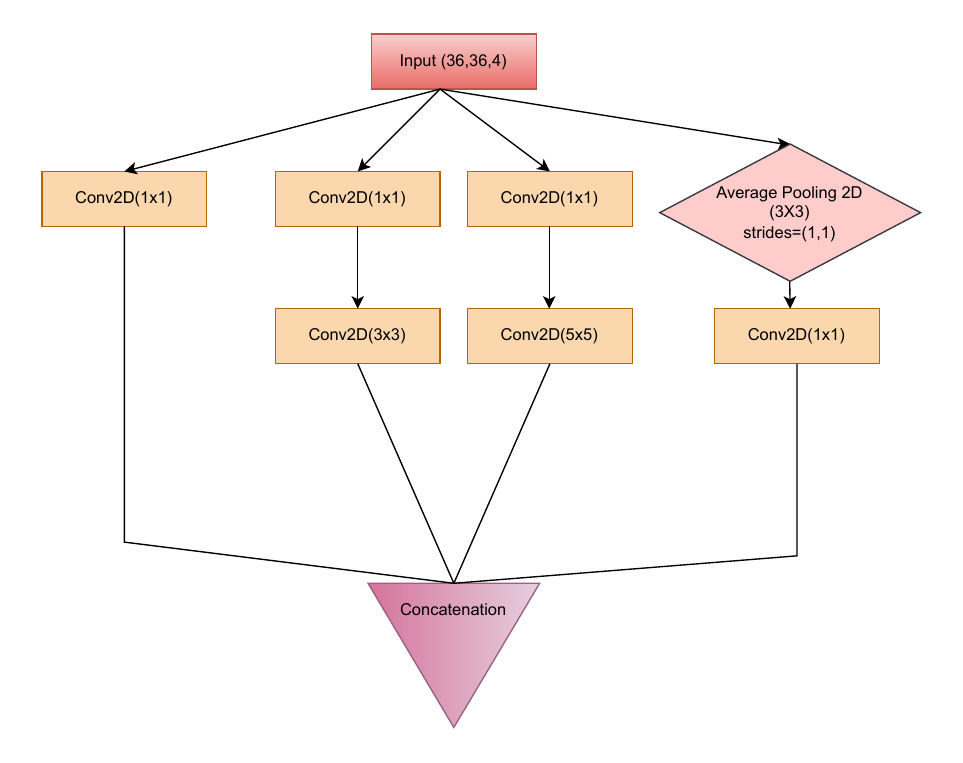}
  \caption
  {Inception module used in this study to estimate the photo-$z$. The input layer has size (36,36,4). Conv2D is a two-dimensional convolution layer, with the kernel size specified in brackets. Average Pooling 2D is a two-dimensional pooling operation that uses a kernel of size \(3 \times 3\). Each operation is represented using boxes of distinct colors. Concatenation is the combined feature maps from parallel convolutions.}
  \label{fig: Inception module}
\end{figure}

In this section, we discuss a CNN architecture referred to as "inception" that we employ in photo-$z$ prediction. Increasing the depth of the network improves the performance at a cost of high computation time. Inception, a deep CNN architecture developed for image detection and classification tasks. It was first used in GoogLeNet architecture \citep{szegedy2015}. The smaller number of weights, and biases compared to its predecessors 
reduces the computation time and makes it appropriate for big-data handling. There are various versions of Inception available, differing by the use of regularization, reduction of overfitting when the training sample is limited, and the inclusion of 
additional DL architectures such as Resnet \citep{Kaiming2015} in the Inception module. In our model, we used the Inception v1 module illustrated 
in Fig. \ref{fig: Inception module}.  We also tested other versions and we found that Inception v1 gives the best performance for our dataset. 

Choosing the right kernel is very important for feature extraction. On the one hand, a larger kernel is suitable for images where the information is distributed globally. However, a smaller-sized kernel is good for locally distributed information extraction. This selection becomes highly important as the galaxies in our sample have a wide range of apparent sizes, and hence our dataset contains both kinds of distributions. We used both larger and smaller-sized kernels in the Inception module. This is shown in Fig. \ref{fig: Inception module}.

The most straightforward way of improving the performance of deep neural networks is by increasing the complexity. This includes both increasing the depth (number of layers) of the network and its width: the number of units at each layer. The inception module uses parallel convolution operation with multiple filter sizes. Inception v1 uses $1\times1$, $3\times3$, and $5\times5$ spatial filters. 

The concatenation process combines all the feature maps from the parallel convolutions. This merging takes place along the channel axis (depth-wise concatenation) which enables the network to efficiently process and extract information from complex visual data. Based on prior research in photo-z estimation, such as  \cite{Henghes2022, Rui2022}, and the performance of our model, we decided to include Inception in our framework.

\subsection{Metrics}
\label{statistics}
In this Section, we discuss 
the metrics to evaluate the performance of our model and photo-zs. Here we discriminate the two types of metrics into those which are evaluated and optimized when building the model (during training and validation), and those computed a posteriori to quantify the behavior of the photo-zs.
For the former, we mainly used three metrics: mean squared error (MSE), mean absolute error (MAE), and the $R$ squared error ($R^2$). 
The MSE and MAE are defined as follows: 

\begin{equation}
    \text{MSE} = \frac{1}{n} \sum_{i=1}^{n} (z_i - \hat{z}_i)^2 ,
\end{equation}

\begin{equation}
    \text{MAE} = \frac{1}{n} \sum_{i=1}^{n} |z_i - \hat{z}_i|. 
\end{equation}
Here, $n$ is the number of samples used for training, $z_i$ is the predicted value and $\hat{z}_i$ is the true value. 
The \(R^2\) is a statistical measure that shows how well a model predicts the outcome in a regression analysis. It is the proportion between the variance explained by the model and the total variance:
\begin{equation}
    R^2 = 1 - \frac{\sum_{i=1}^{n} (z_i - \hat{z}_i)^2}{\sum_{i=1}^{n} (z_i - \bar{z})^2}.
\end{equation}
Here, $\Bar{z}$ is the mean of the spectroscopic redshift values from the GAMA$\times$KiDS crossmatch. For an ideal case, the value of this metric would be unity.

The MSE is more sensitive to outliers than MAE. Due to the squaring part, MSE puts more weight on larger errors. MAE, on the other hand, treats all errors with equal importance, which can be advantageous when the dataset contains extreme values or noisy observations -- which is common for galaxy images.  
Therefore, MSE is more appropriate for learning outliers, and MAE is better for ignoring them. 
To include the benefits of both these loss functions, we use the \cite{Huber1964} loss function, similarly as in the previous KiDS analysis by \cite{Rui2022}. This function is quadratic when the absolute error
is small, and linear when the absolute error exceeds a threshold $\delta$. This makes it robust to outliers because the impact of large errors is reduced compared to using a purely quadratic loss function such as MSE. 
The Huber loss is defined as
\begin{equation}
    {L}(z, \hat{z}) = 
    \begin{cases}
        \frac{1}{2}(z - \hat{z})^2, & \text{if } |z - \hat{z}| \leq \delta, \\
        \delta(|z - \hat{z}| - \frac{\delta}{2}), & \text{otherwise}.
    \end{cases}
\end{equation}

A very low $\delta$ value means that the transition region between the quadratic and linear parts of the loss function is extremely narrow. As a result, even points that are not true outliers but are slightly distant from the predicted values may significantly impact the loss. This can lead to overfitting to outliers. A high value ($\delta> 0.01$ for our case) 
makes the loss function less sensitive to outliers. 
These will affect the model generalization. We tried a range of values between $10^{-5}$ and 0.01, and 
found that $\delta = 0.001$ shows the lowest values for Huber loss, MSE, and MAE. Thus we chose this value of $\delta$ for our models. 

We evaluated the resulting photo-z performance using the following standard statistics:
\begin{itemize}
  \item bias, 
  \begin{equation}
       \delta z= z_\mathrm{phot} - z_\mathrm{spec} ;
  \end{equation}
 
  \item normalized (rescaled) bias, 
  \begin{equation}
       \Delta z= \frac {\delta z}{1+z_\mathrm{spec}} ;
  \end{equation}
  \item standard deviation of normalized bias, $\sigma_{\Delta z}$; %SD(normdz)
  \item scaled median absolute deviation (SMAD) of $\Delta z$, %normdz, SMAD(normdz), 
  where 
\begin{equation}
    \text{SMAD}(x) = 1.4826 \times \text{median}(\lvert x - \text{median}(x) \rvert).
\end{equation}
\end{itemize}
The first two of these metrics quantify the average residuals of the photo-zs from the true value, i.e. their statistical accuracy. The two others measure the scatter, i.e. statistical precision. The factor 1.4826 in the SMAD definition allows it to converge to one standard deviation for the Gaussian.

\section{Photometric redshift model}
\label{model}

Here, we explain the photometric redshift model,  Hybrid-z, used in this study. It incorporates two types of input: galaxy images and magnitudes.

The model was trained using a dataset comprising  
$173k\times4$ KiDS galaxy images from the GAMA equatorial fields and with data augmentation techniques applied, as outlined in Sect. \ref{data augmentation}. Together with the 4-band images, also 9-band magnitudes of the same galaxies were used. Then the model was validated with 26k galaxy samples and tested on an additional set of 26k galaxies also from GAMA equatorial $\times$ KiDS. Finally, we estimated photo-zs for all the KiDS-DR4 Bright sample galaxies. 

We used the Rectified linear unit \citep[Relu,][]{Jentzen2023} as the activation function in all the layers except in the output one where the sigmoid function {(i.e., logistic curve)} is used, which enforces all the predictions to lie in the range $0<z_\mathrm{phot}<1$. {For the layers other than the last (output) one,} we also tried other activation functions such as leaky Relu \citep{Jentzen2023}, peaky Relu, softmax, swish, and tanh{, but Relu gave the best performance.} 

For the output layer, the sigmoid function gives the best results in our redshift range where practically all the galaxies have $z<1$ due to the $r<20$ mag flux limit. For instance, in the entire GAMA spectroscopic sample, objects with $z>1$ constitute less than 0.1\% of the total and these are typically very bright and rare AGNs. For such sparse sources, empirical methods such as ours would not be able to render reliable redshift predictions unless some special approach to anomaly handling is taken. We therefore sacrificed a very small number of objects that lie at true $z>1$ to have $z_\mathrm{phot}<1$ in order to avoid a gross redshift overestimation for others, which could happen if the model had more freedom. The latter is for instance the case for ANNz2, where some of the photo-zs from \citetalias{Bilicki2021} are predicted significantly above unity. {Similar logic applies to photo-$z$s with non-physical predictions of $z<0$ which are equally avoided in our model, while were present in the ANNz2 results. We note, however, that both cases were rare already in \citetalias{Bilicki2021}. In the clean KiDS-Bright sample, there were 82 galaxies with $z_{ANNz2}>1$ and  444 with $z_{ANNz2}<0$.}

\begin{figure}
  \centering
  \includegraphics[width=0.4\textwidth]{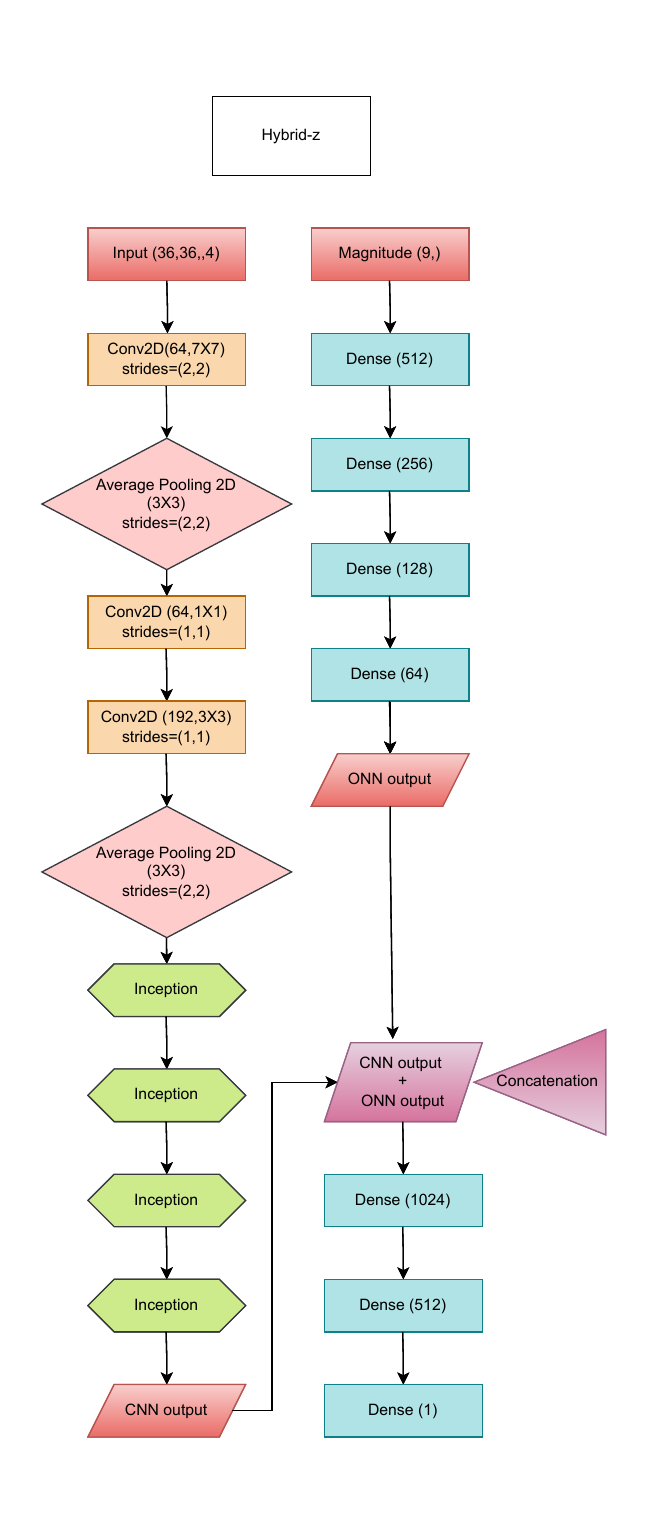}
  \caption{Architecture of the Hybrid-z model, which employs four-band KiDS galaxy images together with nine-band KiDS+VIKING magnitudes for photo-z derivation. For the CNN part (left-hand side), the symbols used are as in Fig. \ref{fig: Inception module}, while inception modules are represented as hexagons. In the fully connected part (right-hand side), dense layers are given as blue rectangles, and the number of neurons is shown in parentheses.}
  \label{combined model fig}
\end{figure}

The architecture of the Hybrid-z model is shown in Fig. \ref{combined model fig}. The left-hand side is the CNN part, where images are processed, while to the right we have the ONN section using magnitudes. For CNNs, the input is $36 \times 36$ pixels ($7.2''\times 7.2''$) galaxy cutouts in four optical bands. When training the network, we used the Adam optimizer \citep{Kingma2014}. One of its key features is the adaptive learning rate, updated during iteration based on previous steps.  $10^{-4}$ is selected as the initial learning rate after various iterative tests with it from $[10^{-5},10^{-2}]$. The loss function is the Huber loss with the $\delta$ hyperparameter value of $10^{-3}$ as mentioned in Sec.~\ref{statistics}. During training, we calculate MAE and MSE in each epoch. When the epoch progresses, their values decrease. 

The features extracted by the convolutional and average pooling layers are passed through the Inception module before reaching the final dense layers. We found that the performance of the model improves with the inclusion of Inception, with notable improvement found in the reduction of the loss function value. We conducted tests with different numbers of Inception modules, ranging from three to five. Our findings suggest that using four Inception modules with varying numbers of filters  
is optimal for our data. The first inception module receives input from the Average Pooling 2D layer, while each subsequent inception module takes its input from the output of the preceding inception module. The initial inception modules extract low-level features and the final inception layers capture the global patterns in the image. {Padding is applied before the convolutional and pooling operations for each Conv2D and AveragePooling2D layer using the  `same' padding \citep{Kaiming2015}. This ensures that the spatial dimensions (height and width) of the output feature maps are maintained as much as possible relative to the input. When the kernel is applied to the input, padding is added around the edges as necessary to keep the spatial dimensions consistent after each convolution and pooling operation.}

\begin{figure}
 \centering
  
  \includegraphics[width=.5\textwidth]{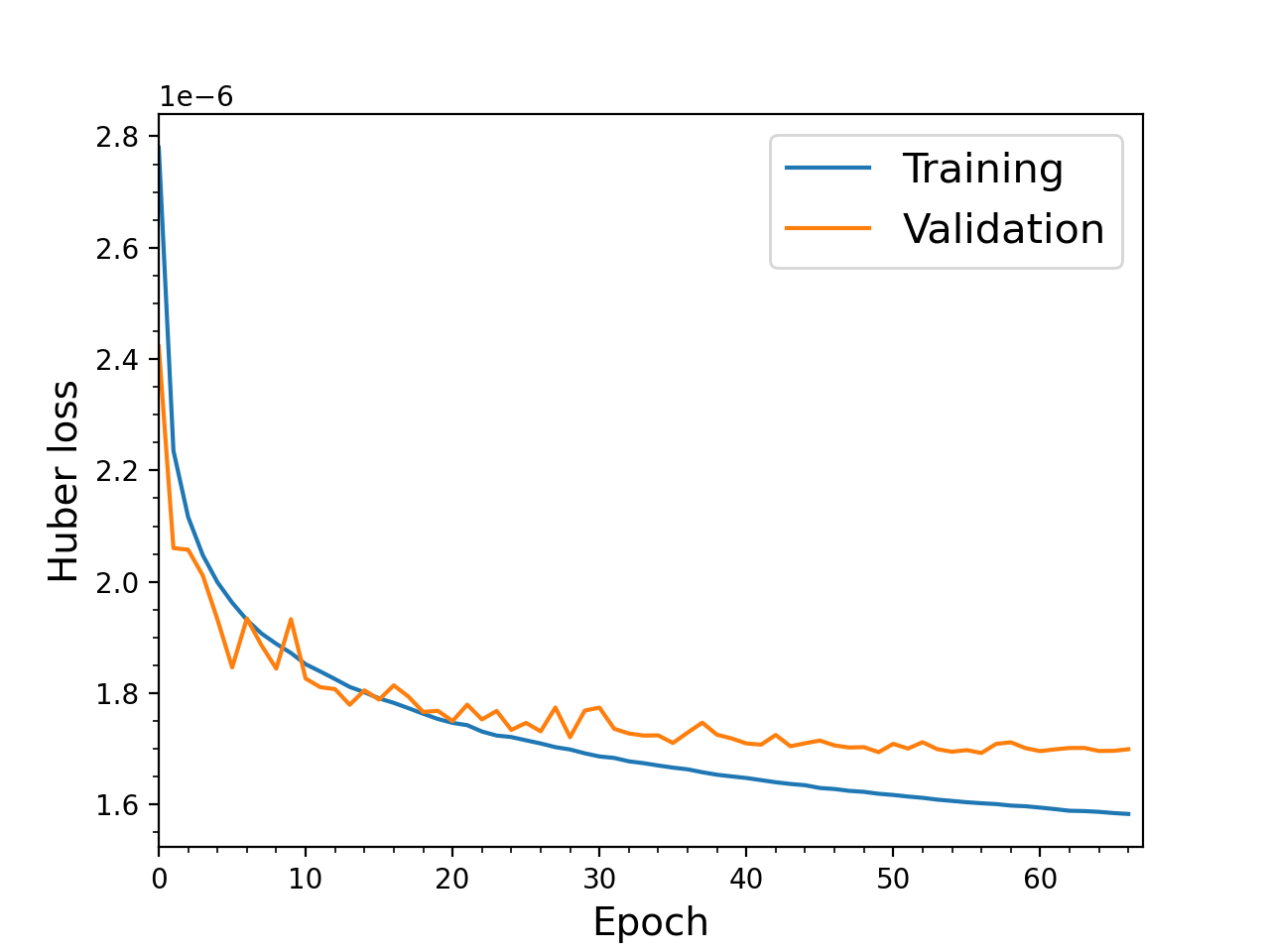}

  \caption{Performance of the Huber loss function during the training and validation stage of the Hybrid-z. }

\label{metrics_combined model}
\end{figure}

As in \cite{Rui2022} and \cite{Henghes2022}, for our photo-z model, we combined two different types of networks, "shallow" (fully connected) ANN, that we will denote as ordinary neural networks (ONNs) and CNN, hence the name Hybrid-z. {There are $\sim$13.8 million trainable parameters in our model.} An important aspect of the Hybrid-z model is the concatenation step. The nine-band magnitudes of galaxies are the additional information for the network. These are
processed by dense layers. The flattened feature map and ONN output are combined via depth-wise concatenation. {ONN output has 64 features. However, when we include images, the number of features increases to 12,064.}
Then this concatenated information is passed through final layers for the prediction of photo-zs.
{This significant increase in features for final dense layers after concatenation highlights the substantial contribution of the image data. }

In Fig.~\ref{metrics_combined model}, the performance of the Huber loss function for the Hybrid-z is shown. The addition of supplementary information enhances the model's performance with respect to using only 4-band images. We find that for this model, we can achieve \(R^2\) value greater than 0.93. 
{We used early stopping criteria \citep{Prechelt1996} for determining the number of epochs. The threshold for considering an improvement is zero, meaning that any decrease in the validation loss would count as an improvement. Early stopping will be triggered after ten consecutive epochs where the validation loss does not improve and the training will automatically stop.}

\section{Results and discussion}
\label{result}

Here we present the results of applying the model described in the previous Section to the KiDS-Bright data discussed in Sec.~\ref{Data}. We 
benchmark our findings against the results obtained previously in \citetalias{Bilicki2021} using `shallow' neural-network software ANNz2. We then analyze the Hybrid-z results for the test sample in more detail. Finally, we apply our best model to the entire KiDS-Bright DR4 photometric sample and discuss the properties of the resulting photo-zs. To reiterate, our training and testing samples with redshift labels have been selected from GAMA equatorial data.

\subsection{Hybrid-z performance on test data}
\label{Test sample}

\begin{table*}
\centering
\begin{threeparttable}

\caption{Statistics of photometric redshift performance obtained for the KiDS-GAMA test sample. Angle brackets $\langle\cdot\rangle$ indicate the mean.} 
\label{table}
\small
\centering

\begin{tabular}{l c c l c c c c c c }
\hline\hline\\

Sample & Size \tnote{1}  
& {$\langle z_\mathrm{spec}\rangle$} & Photo-$z$ model & \multicolumn{1}{c}{$\langle z_\mathrm{phot} \rangle$} & \multicolumn{1}{c}{ $\langle \delta z\rangle$} & $\langle \Delta z\rangle$ & \multicolumn{1}{c}{$\sigma_{\Delta z}$} & \multicolumn{1}{c}{SMAD$(\Delta z)$} \\
\\
\hline\hline\\

      Test data \tnote{2} &26,035 & 0.230 
        
       &  ANNz2 (9-band) &0.230& -0.0002 & 0.0005 & 0.0254 &0.0180\\

        &&&       Hybrid-z (this work)&0.230&-0.0003 &0.0002 &0.0203 &0.0145 \\
        
        \hline\\

        Clean data \tnote{3}&20,965&0.232&ANNz2 (9-band) &0.232& 0.00008 &0.0006&0.0248&0.0178\\
        &&&         Hybrid-z (this work) &0.232&-0.0002&0.0003&0.0197&0.0142\\
        \hline\\

       Red galaxies \tnote{4} &11,062&0.240 & ANNz2 (9-band)  &0.239& -0.0006 & 0.00005  &0.0198 &0.0155 \\
       
       &&& Hybrid-z (this work) &0.239& -0.0006 &0.0002&0.0168 &0.0129 \\
        \hline\\

        Blue galaxies \tnote{4} &12,096&0.212& ANNz2 (9-band) &0.213&0.0007&0.0012 &0.0269& 0.0197\\

        &&& Hybrid-z (this work) &0.212& 0.000001 &0.0004 &0.0212&0.0154

\\
\hline 
\hline

\end{tabular}

\begin{tablenotes}
\item[1] Number of galaxies in the sample.
\item[2] General test data from the KiDS-DR4 Bright sample crossmatched with GAMA in the equatorial fields, including both clean and contaminated galaxies. 
\item[3] Galaxies from the test sample which have the flag masked=0 and are therefore free of any known artifacts \citepalias{Bilicki2021}.

\item[4] Galaxies separated into red and blue based on their position on the absolute $r$-band magnitude - rest-frame $u-g$ color diagram, see \citetalias{Bilicki2021} for details.

\end{tablenotes}

\end{threeparttable}
\end{table*}

\begin{figure}
  \centering
  \subfloat{\includegraphics[width=0.9\columnwidth]{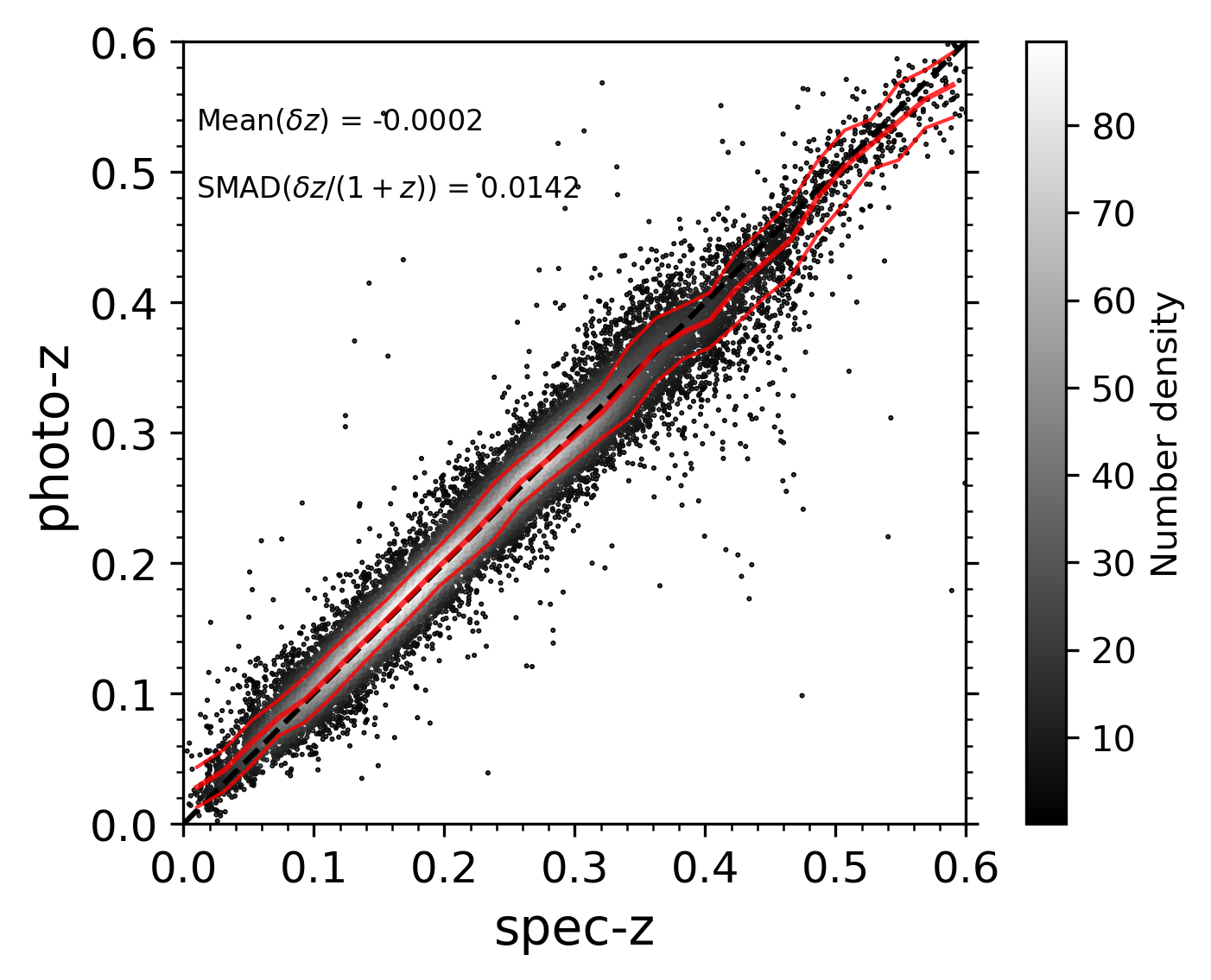}}\\
  
   \caption{Comparison of the true spectroscopic redshifts with the photometric ones derived with our framework, Hybrid-z, using four-band KiDS images and nine-band KiDS+VIKING magnitudes. The test data were taken from a subset of GAMA crossmatched with KiDS. The thick red line depicts the median while the thin lines show the scatter (SMAD) around the median.}
   
  \label{result_combined1}
\end{figure}

\begin{figure*}
       \centering
       
       \includegraphics[width=1.5\columnwidth]
       {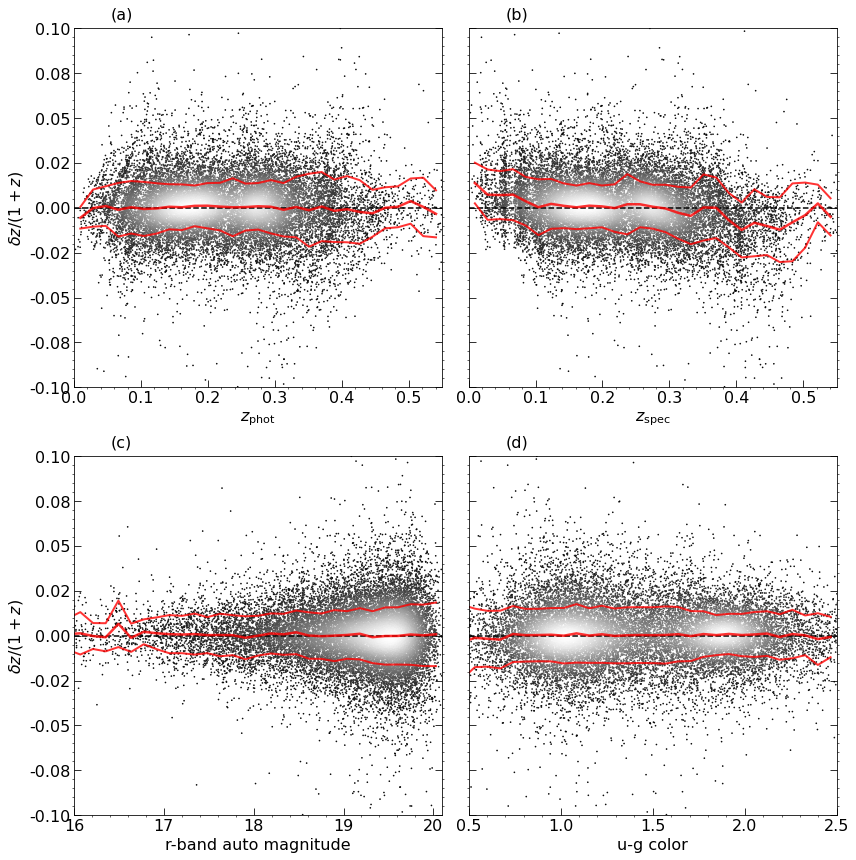}
      \caption{Photometric redshift errors of Hybrid-z as a function of four quantities, from left to right, (a) photometric redshift as derived in this work; (b) true spectroscopic redshift from the clean sample; (c) apparent $r$-band magnitude; and (d) observed $u-g$ color. The thick solid red line represents the running median, while the thin red lines illustrate the scatter (SMAD), based on a blind test sample derived from GAMA.}
  \label{normdz}
\end{figure*}

The comparison of the main statistics of 
Hybrid-z versus ANNz2 results are provided in Table~\ref{table}. Our new model employing jointly optical imaging and KiDS+VIKING magnitudes performs clearly better than ANNz2 which used 9-band magnitudes. In the first block of the Table, we provide statistics of a general test sample, which consists of GAMA galaxies crossmatched with KiDS-Bright including both `clean' (masked=0) and `contaminated' (masked=1) data, where the `masked' flag was derived in \citetalias{Bilicki2021} based on the KiDS-internal `MASK' bit-wise column. The compared methods have comparable mean residuals -- that is, they are similarly accurate -- but Hybrid-z outperforms ANNz2 in its photo-$z$ precision, quantified as the scatter in $\Delta z$. Both in terms of standard deviation and SMAD, our combined CNN+ONN approach performs better, be it for the overall or clean (masked=0) test data. Hybrid-z provides about 20\% smaller scatter than the ANNz2-based derivations of \citetalias{Bilicki2021}, where 9-band magnitudes were used. In the case of the fiducial clean data that are recommended for science, we achieve SMAD$(\Delta z)\simeq 0.014$, as compared to 0.018 in \citetalias{Bilicki2021}. Our result is at the same level as that by \cite{Treyer2023} where CNNs were used for photo-$z$s based on SDSS images at similar depths as in our case. 

To further evaluate the performance and stability of Hybrid-z, we used the $k$-fold cross-validation technique. We employ $k=5$ folds, and the dataset was systematically divided into five distinct subsets. In each iteration of the cross-validation process, one fold was reserved for validation, another for testing, and the remaining three folds were used for training the model. This procedure was repeated five times, each fold taking a turn as the validation and testing set. This ensures that each data point is used for validation, testing, and training, providing a comprehensive evaluation of the model's performance across different subsets of data. The photo-$z$ statistics for each of the folds were very similar, we therefore quote in the tables numbers from only one such runs.

We visualize the performance of the Hybrid-z model in Figs.~\ref{result_combined1} and \ref{normdz}, where we limit the range of redshift shown to $z<0.6$ as beyond that value there are practically no galaxies in our samples (for instance, there are only 18 objects with $z>0.6$ in the test data). Figure \ref{result_combined1} directly compares the true redshifts from the test set with our photo-z predictions.  The running median (depicted by the thick red line) closely follows the diagonal for most of the $0<z<0.6$ range, deviating %slightly 
only at the lowest and highest redshifts. This behavior is typical for ML photo-$z$ derivations, which are unbiased as a function of {photometric}, but not {spectroscopic} redshifts. This is further illustrated in 
panels (a) and (b) of Fig.~\ref{normdz},
where photo-z residuals (rescaled by $1+z$ as typically done) are plotted as a function of respectively predicted photo-$z$ (panel a) and of the true spectroscopic redshift (panel b).

In panels (c) and (d) of Fig.~\ref{normdz} we illustrate the behavior of photo-$z$ errors of our Hybrid-z model as a function of two observables: the $r$-band apparent magnitude and the observed $u-g$ color. The former is shown for the `auto' flux measurement, which approximates the total light of a galaxy and was used by \citetalias{Bilicki2021} to select the flux-limited KiDS-Bright sample. We observe slightly growing scatter (thin red lines indicating SMAD in Fig.~\ref{normdz} (c)) 
at the faintest end 
of the sample ($r\lesssim20$). Otherwise, photo-$z$s are stable as a function of magnitude, with practically zero bias at most of the range. The $u-g$ color shown in panel (d) 
can be used to split galaxies into red and blue populations (similarly as $u-r$, \citealt{Strateva2001}). Indeed, we see the bimodality in the plot, with bluer galaxies to the left and redder to the right. As expected for photo-$z$s, the former has a wider scatter with respect to the true value than the latter. The difference is, however, not substantial, and again the median biases for the whole range of the $u-g$ color are very close to zero.

We further quantify the performance of Hybrid-z for blue and red galaxies in the two bottom blocks of Table \ref{table}. Here, for a direct comparison between this work and \citetalias{Bilicki2021}, the selection of the two galaxy types was based on their intrinsic rather than observed properties, namely via their positioning on the absolute $r$-band magnitude versus rest-frame $u-g$ color diagram\footnote{More specifically, we crossmatched our sample with galaxies labeled as red and blue in KiDS-Bright DR4.} as discussed in \citetalias{Bilicki2021}.
Similarly as for the full sample, also for these subsets, Hybrid-z provides considerable improvement in precision over ANNz2. Interestingly, this {reduction in SMAD($\Delta z$)} is greater for blue galaxies (by $\sim22\%$) than for red ones ($\sim17\%$). This is consistent with the fact that, as compared to the approaches relying solely on summary measurements such as magnitudes, the  DL-based methods, which directly extract features from images to estimate redshifts, are expected to show larger improvement for galaxies with intricate morphologies such as spirals, that typically have bluer colors \citep[e.g.,][]{Schuldt2021,Treyer2023}. In contrast, for red galaxies -- typically ellipticals with fewer features in images -- the reduction in scatter for Hybrid-z is smaller than for the overall sample, although we would such as to emphasize that it still does perform better than ANNz2 for the same galaxy selection. In addition, it is worth noting that SMAD($\Delta z$) of 0.0154 for {blue} galaxies as obtained by Hybrid-z is even smaller than the same statistic of ANNz2 for {red} ones (cf. blocks \#3 and \#4 of Table~\ref{table}).

\begin{figure}
    \centering
    \includegraphics[width=0.99\columnwidth]{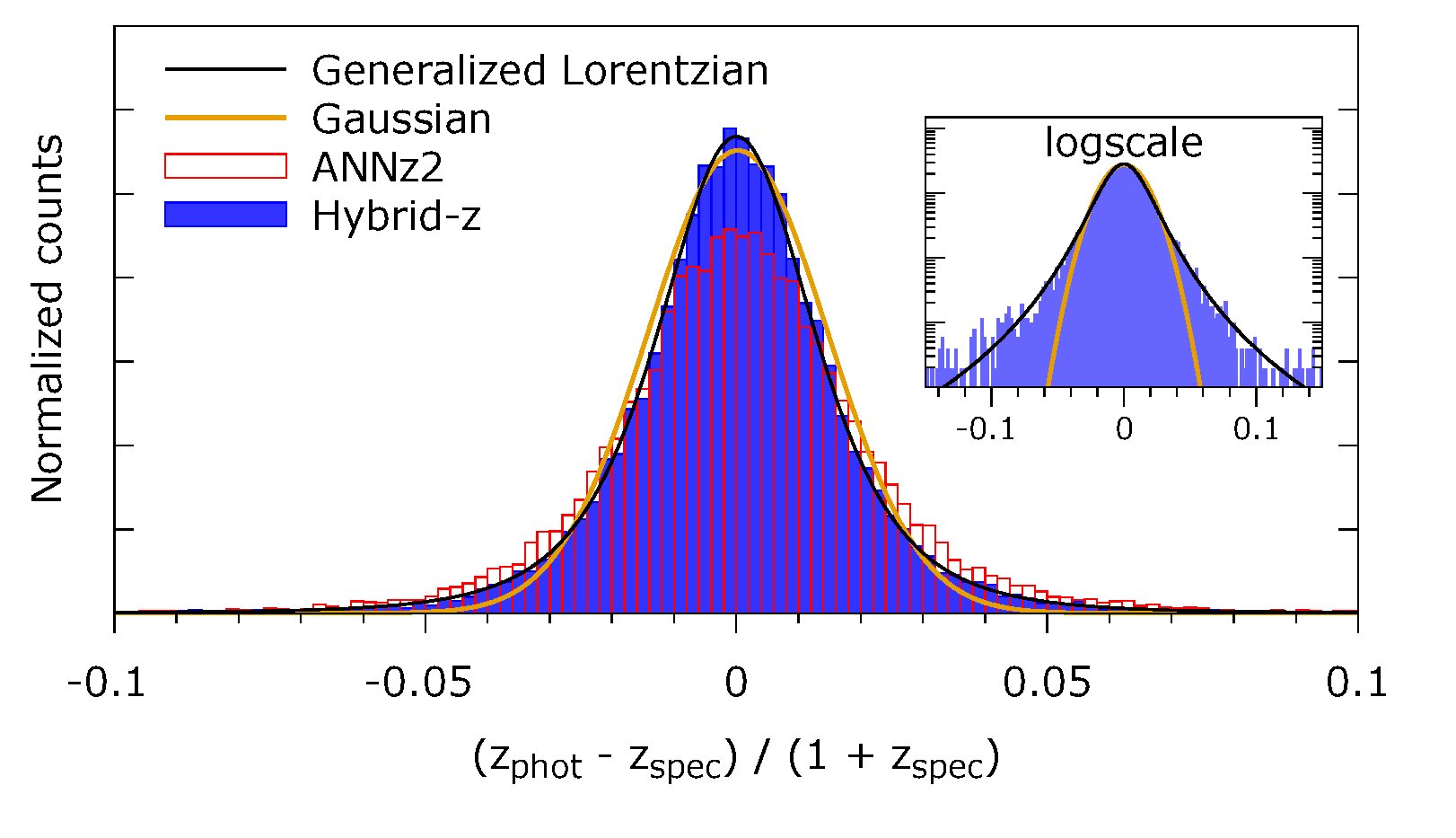}
    \caption{Comparison of photometric redshift error distributions between Hybrid-z (blue, this work) and ANNz2 (\citetalias{Bilicki2021}, red) shown as normalized counts based on the test sample results. Solid lines illustrate the best-fit Gaussian (with $\sigma=0.0145$ and $\mu=2.56\times10^{-4}$, yellow) and generalized Lorentzian (Eq.~\ref{Eq:gen-Lorentz}, with $a=2.332$ and $s=0.0116$, black). The inset compares the histogram of Hybrid-z photo-z residuals with the fit models in logscale of the $y$-axis.} 
    \label{fig:Delta z}
\end{figure}

The final comparison between ANNz2 and Hybrid-z performance is provided in Fig.~\ref{fig:Delta z}, where we show the distributions of photo-$z$ errors $\Delta z$ as calibrated on the GAMA test data. As discussed in \citetalias{Bilicki2021}, following earlier work \citep[e.g.,][]{Bilicki2014}, such a distribution presents non-Gaussian features in the wings and is therefore better fit with a "generalized Lorentzian" of the form\footnote{It is interesting to note that also spectroscopic redshifts may present Lorentizan rather than Gaussian uncertainties, e.g., \cite{Yu2024}.}
\begin{equation} \label{Eq:gen-Lorentz}
 N(\Delta z) \propto \left(1+\frac{\Delta z^2}{2 a s^2}\right)^{-a},  
\end{equation}
where the mean bias is assumed to be negligible (but see \citealt{Hang2021} on including non-zero mean). For our Hybrid-z model in the test sample, the best-fitting values are $a=2.332$ and $s=0.0116$. If we instead fit a Gaussian, this time allowing the mean to depart from 0, we get best-fit values of $\sigma=0.0145$ and $\mu=2.56\times10^{-4}$. As visible in Fig.~\ref{fig:Delta z}, the Gaussian is a worse fit to the residuals than Eq.~\eqref{Eq:gen-Lorentz}. In the same plot, we also show the histogram for ANNz2 residuals as derived by \citetalias{Bilicki2021}, which is clearly broader. Quantitatively, the above numbers can be compared with best-fit $\{a,s\}_\mathrm{B21}=\{2.613,0.0149\}$ and $\sigma_\mathrm{B21}=0.0180$. For the Gaussian, the reduction in scatter of Hybrid-z is by 24\%. For the modified Lorentzian, the width is encoded in the $s$ parameter while $a$ quantifies the extent of the wings; we see a reduction in both for Hybrid-z in comparison to ANNz2.

To summarize this subsection, the Hybrid-z model using CNNs on KiDS optical images together with ONNs on KiDS+VIKING magnitudes 
provides considerably better results than the previous 9-band ANNz2 derivations of \citetalias{Bilicki2021}. The improvement from ANNz2 to Hybrid-z is at the level of 20\% reduction in photo-$z$ scatter (better precision) while maintaining almost zero bias (very good accuracy). Thanks to our new model, we observe smaller photo-$z$ errors for various subsamples of test data, and most remarkably for blue galaxies, where SMAD($\Delta z$) is lowered by about 22\% from what was obtained in \citetalias{Bilicki2021}. As such, these are state-of-the-art results for the KiDS-Bright galaxies and in line with independent derivations for SDSS at the same depth \citep{Treyer2023}.

\subsection{Validation of Hybrid-z across external spectroscopic samples}
\label{Sec: external spec-z}

\begin{table*}
\centering
\begin{threeparttable}

\caption{Statistics of photometric redshift performance for KiDS-Bright crossmatched with various spectroscopic samples.} 
\label{table:extra-zspec}
\small

\begin{tabular}{l r c l c c c c c c }
\hline\hline\\

Sample & \multicolumn{1}{c}{Size \tnote{1}} & {$\langle z_\mathrm{spec}\rangle$} & Photo-$z$ model & \multicolumn{1}{c}{$\langle z_\mathrm{phot} \rangle$} & \multicolumn{1}{c}{ $\langle \delta z\rangle$} & $\langle \Delta z\rangle$ & \multicolumn{1}{c}{$\sigma_{\Delta z}$} & \multicolumn{1}{c}{SMAD$(\Delta z)$} \\
\\
\hline\hline\\

      GAMA Equatorial \tnote{2} & 145,493 & 0.229 &  ANNz2 & 0.229 & 0.0005 & 0.0009 & 0.0237 & 0.0178\\
        &&&       Hybrid-z (this work) & 0.224 & -0.0044 & -0.0031 & 0.0187 & 0.0144 \\
        
        \hline\\

        2dFGRS \tnote{3} & 53,179 & 0.119 & ANNz2 & 0.122 & 0.0022 & 0.0022 & 0.0238 & 0.0158\\
        &&&         Hybrid-z (this work) & 0.120 & 0.0004 & 0.0006 & 0.0167 & 0.0123\\
        \hline\\
              
        SDSS DR16 \tnote{4} & 43,581 & 0.221 & ANNz2  & 0.220 & -0.0011 & -0.0002 & 0.0221 & 0.0157\\
        &&& Hybrid-z (this work) & 0.217 & -0.0034 & -0.0023 & 0.0158 & 0.0116 \\
        
        \hline\\
          
       GAMA G23 \tnote{5} & 34,941 & 0.215 & ANNz2 & 0.217 & 0.0021 & 0.0022  & 0.0244 & 0.0166 \\
       
       &&& Hybrid-z (this work) & 0.213 & -0.0027 & -0.0017 & 0.0184 & 0.0137 \\
        \hline\\
                         
        2dFLenS \tnote{6} & 22,128 & 0.251 & ANNz2 & 0.251 & -0.0007 & 0.0005 & 0.0365 & 0.0167\\
                
        &&& Hybrid-z (this work) & 0.247 & -0.0042 & -0.0025 & 0.0224 & 0.0133\\

\hline 
\hline

\end{tabular}

\begin{tablenotes}
\item[1] Number of galaxies in the crossmatched sample.
\item[2] Galaxy And Mass Assembly in the equatorial fields \citep{Driver2022}, used for training both models. 
\item[3] 2-degree Field Galaxy Redshift Survey \citep{2dFGRS}.
\item[4] Sloan Digital Sky Survey Data Release 16 \citep{SDSS-DR16}.
\item[5] GAMA in the southern G23 field \citep{Driver2022}, disjoint with the training set.
\item[6] 2-degree Field Lensing Survey \citep{2dFLenS}.
\end{tablenotes}

\end{threeparttable}
\end{table*}

In addition to the standard evaluation of Hybrid-z performance from Sec.~\ref{Test sample} done on test samples statistically consistent with the training data (as both were randomly sampled from input GAMA equatorial catalogs), we have done a further validation of our photo-$z$s.
Still using GAMA-Equatorial for training, we checked Hybrid-z predictions for a number of spectroscopic samples overlapping with the KiDS DR4 area and covering appropriate magnitude ranges and sky areas to provide sufficient statistics. These include GAMA Equatorial and G23 fields \citep{Driver2022}, 2dFGRS \citep{2dFGRS}, SDSS DR16 \citep{SDSS-DR16}, and 2dFLenS \citep{2dFLenS}.
Among these, GAMA G23 and 2dFLenS are disjoint with our GAMA-Equatorial training set, as they cover only the southern patch of KiDS (at $\delta<-25^\circ$)\footnote{A very small fraction ($\sim1\%$) of 2dFLenS is located in KiDS-N, but the overlap with GAMA-Equatorial fields is negligible ($\sim200$ sources).}. The surveys SDSS and 2dFGRS do have common objects with GAMA in its equatorial patches. However, they extend to the much wider KiDS area, so their overlap with GAMA-eq is also small.

In Table \ref{table:extra-zspec} we present the statistics derived from crossmatching the above-mentioned datasets with the full KiDS-Bright `clean' sample. For each spectroscopic dataset, we compare the performance of Hybrid-z with previous ANNz2 results from \citetalias{Bilicki2021}. For each of these test sets, Hybrid-z performs considerably better in terms of scatter than ANNz2, while generally the former displays larger bias than the latter. {This indicates that our new model loses some of the accuracy that ANNz2 had while gaining in precision (typical ML trade-off). We note, however, that in all the inspected cases, for Hybrid-z the mean biases in $\delta z$ or $\Delta z$ are much smaller than the scatter, meaning that the model is still highly accurate in its photo-z predictions. In the future, we plan to inspect this further with KiDS DR5 \citep{Wright2024} 9-band imaging and extending the training with, for example, DESI DR1 \citep{DESI-DR1}, to see if we can minimize both precision and accuracy at the same time.}

What is worth noticing is the very small SMAD of Hybrid-z ($\sim0.012$) for the SDSS DR16 crossmatch. This dataset, due to the $r<20$ flux limit of KiDS-Bright, includes galaxies from the SDSS main sample (flux-limited to $r<17.77$), as well as a subset of BOSS LOWZ and CMASS, dominated by luminous red galaxies. This combination leads to the photo-z statistics for such a SDSS $\times$ KiDS-Bright crossmatch being overall better than for the general red galaxies specified in Table \ref{table}. However, the same is not the case for ANNz2, which performs only marginally better for such a mixture of SDSS galaxies than it did for the general red galaxy selection.

Among the datasets included in Table \ref{table:extra-zspec}, GAMA G23 and 2dFLenS are genuinely `blind' test samples, as they are separated on-sky by many degrees from the area where the GAMA-Equatorial training sets are. Photo-$z$ statistics derived from these catalogs should therefore be robust against possible overfitting thanks to their independence from the training data. While one should remember that neither G23 nor 2dFLenS are as complete and flux-limited samples as KiDS-Bright and GAMA-Equatorial, it is reassuring to find that for both of the former Hybrid-z gives considerable improvement over ANNz2, namely reduction in scatter by respectively 17\% and  20\%. 

To summarize, in addition to evaluating the Hybrid-z model on test samples similar to the training data, we validated its photo-$z$ predictions using several independent spectroscopic datasets within the KiDS DR4 survey area. Results show that Hybrid-z outperforms the previous ANNz2 model in terms of scatter across all test datasets, indicating robust performance also on datasets statistically different from the training.

\subsection{Application to the KiDS-DR4 Bright sample}

The Hybrid-z model trained on the full GAMA equatorial data, as discussed above, provides satisfactory results when tested on various overlapping spectroscopic datasets. However, applying it directly to the KiDS-Bright sample introduces artifacts in the final photo-$z$ distribution, which {we believe} are related to the specific properties of the GAMA training set. Namely, GAMA sky coverage of 180 deg$^2$ is small enough that the survey is affected by cosmic variance, manifesting itself by enhanced effects of the cosmic web, such as voids and filaments \citep[e.g.,][]{Eardley2015}. This results in significant `peaks' and `dips' in the GAMA redshift distribution, which should average out for larger sky area. {However, these kind of features are still imprinted} onto our 
photo-$z$ predictions. Namely, training the Hybrid-z model directly on the full GAMA dataset leads to $dN/dz_\mathrm{phot}$ of KiDS-Bright which mimics some of the LSS-related properties of GAMA $dN/dz_\mathrm{spec}$.

\begin{figure}
    \centering
    \includegraphics[width=\columnwidth]{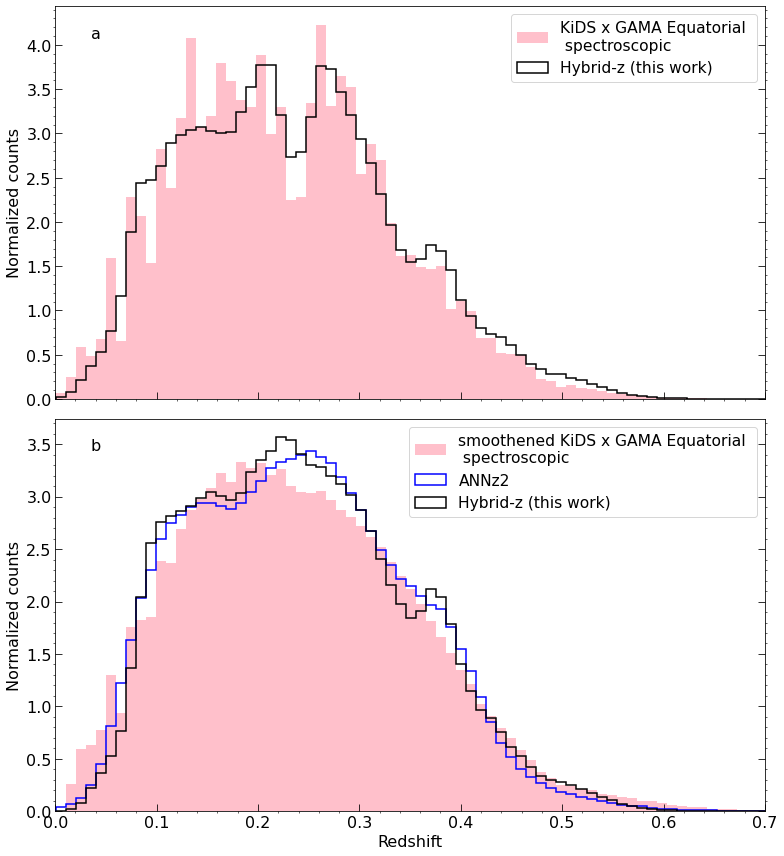}
   \caption{Photo-$z$ distribution of the KiDS-DR4 Bright sample ($r < 20$ mag) as predicted by ANNz2 (blue, \citetalias{Bilicki2021}) compared with photo-$z$ estimates from Hybrid-z (this work) trained on the spec-$z $ distribution of (a) KiDS x GAMA bright sources (original training sample) (b) smoothed KiDS x GAMA bright sources. }

    \label{hist_full}
\end{figure}

As discussed in previous papers \citep[e.g.,][\citetalias{Bilicki2021}]{Bilicki2016}, the redshift distribution of GAMA has strong features, 
and in particular, there is an under-abundance of galaxies at $z_\mathrm{spec}\sim0.25$. This redshift is close to the median of the sample, which is where ML models typically work optimally. If we then train on such a specific distribution, our model becomes biased toward this input $dN/dz$, which results in a `dip' at $z_\mathrm{phot}\sim0.25$ and two peaks below and above this value (see panel (a) of Fig.~\ref{hist_full}). As photo-zs dilute the structures in the radial direction, for a sample covering $\sim1000$ deg$^2$ such as ours, this kind of strong features in its $dN/dz_\mathrm{phot}$ are {unlikely to be} physical but rather originate from `redshift focusing' of the model. {However}, this behavior was not observed in \citetalias{Bilicki2021}, where photo-$z$s were {derived with ANNz2.}

Our interpretation is that {our new model is more sensitive to such strong features in the training data and this needs to be mitigated. One possibility would be to include training sets covering more of the sky, and hence they would be less affected by cosmic variance. This will be possible thanks to, for example, DESI.\footnote{DESI Data Release 1 \citep{DESI-DR1} was released after this work had been completed; hence, we do not include it here.} Another option is to train several models, with for instance different random seeds and/or architectures, and appropriately combine their outputs into the final prediction. Such an approach is implemented in ANNz2, but it uses much less computationally demanding ONNs. Employing a similar framework with DL would be beyond our scope of research. Below we propose another mitigation strategy consisting of appropriately resampling ("smoothing") the available training set.} 
%\sout{the improved sensitivity of our new model reveals these artifacts, indicating that training data specifics now significantly influence model behavior.} 

The second effect we observe is related to the mismatch between the training and target photometric datasets at the faint end. {Namely, GAMA is complete to a brighter magnitude than the KiDS-Bright selection, which will lead to some extrapolation of model predictions at the faint end.} As quantified in the recent paper by \cite{Jalan2024}, the GAMA-equatorial sample becomes considerably incomplete with respect to KiDS-Bright at $r\gtrsim 19.5$ mag. For a supervised ML model such as ours, this leads to extrapolation resulting from the so-called
covariate shift. ML models tend to perform well when the test and training data share a well-matched feature space and distribution of the target quantity. When this distribution shifts or a covariate shift occurs, model performance can be adversely affected \citep{Dharani2019}. In our case, this affects the predictions at the faintest magnitudes of the sample. In particular, as was already the case for ANNz2, many galaxies with $r>19.5$ are assigned $z_\mathrm{phot}>0.35$ where training data is sparser. For Hybrid-z, training directly on the full GAMA dataset introduces additionally a new peak at $z_\mathrm{phot}\sim0.38$, revealing a redshift focusing effect in our model’s predictions as evident from panel (a) of Fig.~\ref{hist_full}). 

In order to solve the above-discussed issues, for the final training of the full-sample photo-zs, we decided to subsample the spectroscopic redshifts from the GAMA equatorial dataset in such a way as to smooth out the peaks and dips originally present. The subsampling was done in such a way to not affect the color-redshift relation but instead provide a `smoother' (more regular) input redshift distribution in GAMA without the strong features discussed above. We create this smoothed subsample of training data by iteratively adjusting the distribution of redshift values to achieve a more uniform histogram. Starting with an initial histogram of true $z$ values, bins with counts that significantly exceed their neighboring bins are identified as spikes, based on a decreasing threshold. In each iteration, if a bin exceeds the count of its neighbors by this threshold, its count is reduced to either the average of its neighbors or its original count, whichever is lower. Randomly selected data points are then drawn from each bin according to the adjusted counts, preserving the dataset's structure but smoothing out extreme values. This iterative process results in a more evenly distributed subsample, which is useful for downstream tasks, by reducing overrepresented regions. Finally, the resulting `KiDS ID' values and their corresponding $z$ values for this subsample are output, with a calculation of the subsample size as a percentage of the original dataset. This gave us an output of 118k galaxies from the GAMA training set (about $66\% $of the original one) and their redshift distribution is shown as red bars in panel (b) in Fig.\ref{hist_full}. Using these smoothed data we retrained the Hybrid-z model and then applied it to the full $r<20$ mag dataset of $\sim1.2$ million KiDS DR4 galaxies. Among these, about 996k have the flag `masked=0' indicating their usefulness for science (see \citetalias{Bilicki2021} for details).

In panel (b) of Fig.\ref{hist_full} we compare the dN/d$z_{phot}$ of the full sample from our model (Hybrid-z) and ANNz2 from \citetalias{Bilicki2021}. We also show the spec-$z$ distribution of the smoothed training sample. Differences in the photo-$z$ distributions are notable between Hybrid-z and ANNz2, particularly at $z \sim 0.24$ and within the redshift range of $(0.3,0.4)$. Otherwise, they are very consistent, despite the fact that ANNz2 redshifts were trained on the full GAMA sample as shown in panel (a), while Hybrid-z used the smoothed GAMA subsample for training, as shown in panel (b). What persists in our derivations is the peak at $z_\mathrm{phot}\sim0.38$, which is present whether we train on original or smoothed GAMA data. {We note that the comparison between Hybrid-z and ANNz2 $dN/dz_\mathrm{phot}$ serves here just as a cross-check, but the aim is not to make them overlapping. While some consistency between the two approaches is expected, as the models use the same input training data and were applied to the same inference sample, direct comparisons between two photo-z approaches should be done with care. For science applications of such data, further redshift calibration is needed to either reproduce the underlying $dN/dz_\mathrm{true}$ or to build a photo-z error model.}

To summarize, the Hybrid-z model, trained on the full GAMA equatorial dataset, performs well on overlapping spectroscopic surveys, but introduces artifacts when applied directly to the KiDS-Bright sample, due to the specific properties of the GAMA training set. To mitigate these problems, a smoothed subsample of the training data was created to achieve a more uniform redshift distribution, which was then used to retrain the model. As a result, we obtain photo-z predictions for the full KiDS-DR4 Bright galaxy sample, which displays improved performance over previous derivations and also gives a generally artifact-free redshift distribution.

We release the photometric redshifts generated with our Hybrid-z model for the entire KiDS-DR4 Bright Sample as a supplement to the original dataset which was accompanying \citetalias{Bilicki2021}. This is available from the KiDS webpage at \url{https://kids.strw.leidenuniv.nl/DR4/brightsample.php}. 

\section{Conclusions and future prospects}
\label{conclusion}
This work presents the first DL photometric redshift (photo-$z$) derivations for the flux-limited KiDS-Bright galaxy sample with the selection threshold 
$r<20$ mag \citep{Bilicki2021}. Previously, photo-$z$s  for this catalog were estimated using "shallow" learning methods, specifically the ANNz2 neural network package \citep{Sadeh2016}. Our new model, Hybrid-z, is built on recent studies, including \cite{Rui2022}, where DL was applied for photo-$z$s in a deeper KiDS galaxy sample, and \cite{Treyer2023}, who used a similar training set as us to obtain CNN-based redshifts for an SDSS-selected catalog with the same flux limit.

We built and tested a DL model for photo-$z$ derivation called Hybrid-z 
that uses four-band KiDS images ($ugri$) processed by CNNs, which are combined with nine-band magnitudes from KiDS+VIKING, processed by an ONN. Rather than simply averaging the outputs of the two networks, we concatenated their outputs before the final three dense layers. This approach was inspired by the previous works of, for example, \cite{Rui2022} and \cite{Henghes2022}, and it yielded significantly improved photo-$z$ performance compared to the previous ANNz2 derivations of \citetalias{Bilicki2021}, where nine-band KiDS+VIKING magnitudes were used. The Hybrid-z model reduces the scatter (SMAD) of $\Delta z$ by 20\% as compared to ANNz2 for the same test samples. This is true for multiple spectroscopic test datasets, of which some can be considered entirely "blind" in terms of being fully disjointed with the training data.
When tested on the fiducial 'clean" KiDS-Bright sample, Hybrid-z achieves a SMAD$(\Delta z)$ of approximately $0.014(1+z)$, representing a clear 
improvement over the prior ANNz2 results giving $\sim0.018(1+z)$ while maintaining the same minimal bias in $\delta z$ of at most a few times $10^{-4}$. 

The Hybrid-z model shows even greater improvement in photo-$z$ precision when we separate out blue galaxies from the KiDS-Bright sample. For these objects, 
Hybrid-z reduces the SMAD of $\Delta z$ by 22\% as compared to ANNz2, resulting in a scatter comparable to that which ANNz2 attained for red galaxies. This of course, means that at the same time, the photo-$z$ performance for red galaxies improves less than on average (by 17\% in SMAD) after adding DL as compared to the ordinary "shallow" networks. This is consistent with expectations, as CNNs can leverage the detailed varied features in blue (typically spiral) galaxies more effectively than in the smoother elliptical red galaxies.
In a flux-limited sample at low redshift, as ours, blue galaxies are more abundant than red galaxies, so such a property of the photo-$z$ model is very useful to improve the overall quality of the derivations. 
These advancements pave the way for more refined astrophysical and cosmological analyses using KiDS-Bright data, such as of the stellar-to-halo-mass relation \citepalias{Bilicki2021} or multi-probe analyses \citep{Dvornik2023}.

Our new photo-$z$ model trained on the GAMA equatorial dataset performs reliably on KiDS overlapping spectroscopic data, but it introduces artifacts in the redshift distribution when directly applied to the KiDS-Bright sample. These artifacts stem from GAMA's limited sky coverage and cosmic variance, which affect the resulting $dN/dz_\mathrm{phot}$ and introduce specific patterns that mimic GAMA characteristics, such as a notable dip at  $z_\mathrm{phot} \sim 0.25$, and extrapolation issues at faint magnitudes and where $r>19.5$ mag. To overcome these limitations, we created a smoothed subsample of the GAMA training data by reducing sharp peaks and dips in its redshift distribution and achieving a more uniform $dN/dz_\mathrm{spec}$ for final model training. This approach mitigates artifacts in the output $dN/dz_\mathrm{phot}$, thus enhancing the reliability of our photo-$z$ estimates. These improved photo-$z$s for the KiDS-DR4 Bright sample ($\sim$1.2M galaxies) are available publicly at \url{https://kids.strw.leidenuniv.nl/DR4/brightsample.php}.

In this work, our DL analysis focused on the four-band KiDS optical images, while the full nine-band KiDS+VIKING photometry was used only in the form of magnitudes. Moving forward, we plan to build an extended DL model incorporating images from all nine bands, from KiDS $u$ to VIKING $K_s$, once the NIR coadds are available; currently, they are being prepared for the 4MOST WAVES target selection \citep{Driver2019}. We anticipate that this expansion will improve CNN-based photo-$z$ precision and will mirror the gains seen from KiDS DR3 $ugri$ \citep{Bilicki2018} to nine-band inclusion in DR4 \citep{Bilicki2021}. We plan to apply such an enhanced Hybrid-z model to the final KiDS Data Release 5 \citep{Wright2024} for state-of-the-art bright-end photometric redshifts. Additional improvement in DR5 is expected thanks to the second $i$-band pass, which deepens effective imaging and should help further refine our photo-$z$ estimates. Last but not least, our approach holds promise for even deeper imaging applications in the forthcoming Legacy Survey of Space and Time \citep{LSST2009}, paving the way for robust  photo-$z$s in future large-scale sky surveys.

\section*{Data availability}
The photometric redshift catalog based on the Hybrid-z DL model, containing redshifts for over 1.2 million galaxies in the KiDS-Bright DR4 sample, is available at the CDS via
anonymous ftp to \url{cdsarc.cds.unistra.fr (130.79.128.5)}
or via \url{https://cdsarc.cds.unistra.fr/viz-bin/cat/J/
A+A/698/A276}.
\begin{acknowledgements}
We would like to thank Elisa Chisari, Rui Li, Nicola Napolitano \& Angus Wright for their valuable comments and suggestions on the manuscript.\\

Based on data products from observations made with ESO Telescopes at the La Silla Paranal Observatory under program IDs 177.A-3016, 177.A-3017 and 177.A-3018, and on data products produced by Target/OmegaCEN, INAF-OACN, INAF-OAPD and the KiDS production team, on behalf of the KiDS consortium. OmegaCEN and the KiDS production team acknowledge support by NOVA and NWO-M grants. Members of INAF-OAPD and INAF-OACN also acknowledge the support from the Department of Physics \& Astronomy of the University of Padova, and of the Department of Physics of Univ. Federico II (Naples).\\

GAMA is a joint European-Australasian project based around a spectroscopic campaign using the Anglo-Australian Telescope. The GAMA input catalog is based on data taken from the Sloan Digital Sky Survey and the UKIRT Infrared Deep Sky Survey. Complementary imaging of the GAMA regions is being obtained by a number of independent survey programs including GALEX MIS, VST KiDS, VISTA VIKING, WISE, Herschel-ATLAS, GMRT, and ASKAP providing UV to radio coverage. GAMA is funded by the STFC (UK), the ARC (Australia), the AAO, and the participating institutions. The GAMA website is \url{http://www.gama-survey.org/}.\\

This work is supported by the Polish National Science Center through grants no. 2020/38/E/ST9/00395, and 2018/31/G/ST9/03388.\\
We have made use of \textsc{TOPCAT} \citep{Taylor2005} and \textsc{STILTS} \citep{Taylor2006} software, as well as of \textsc{python} (\url{www.python.org}), including the packages \textsc{NumPy} \citep{harris2020}, \textsc{SciPy} \citep{Virtanen2020}, and \textsc{Matplotlib} \citep{Hunter2007}.\\
\end{acknowledgements}

\bibliographystyle{aasjournal}
\bibliography{references}

\end{document}